\documentclass [12pt]{article}
\def\maj#1{\ifmmode\mbox{\usefont{U}{msb}{m}{n}#1}\else{\usefont{U}{msb}{m}{n}#1}\fi}
\def\v#1{\mathbf{#1}}
\usepackage{graphics,epsfig,graphicx}
\usepackage{color}

\begin{document}

\title{\textbf{The exciton many-body theory extended to arbitrary
composite bosons}}
\author{M. Combescot$^a$, O. Betbeder-Matibet$^a$ and F. Dubin$^b$
\\ $^a$\small{\textit{Institut des NanoSciences de Paris,}}\\
\small{\textit{Universit\'e Pierre et Marie Curie-Paris 6, 
Universit\'{e} Denis Diderot-Paris 7, CNRS, UMR 7588,}}\\
\small{\textit{Campus Boucicaut, 140 rue de Lourmel, 75015 Paris}}
\\$^b$\small{\textit{Institute for Experimental Physics,}}\\
\small{\textit{University of
Innsbruck, Technikerstr.\ 25, A-6020 Innsbruck, Austria}}} 
\date{}
\maketitle

\begin{abstract}
We have recently constructed a many-body theory for composite excitons, in
which the possible carrier exchanges between $N$ excitons can be treated
exactly through a set of dimensionless ``Pauli scatterings'' between two
excitons. Many-body effects with excitons turn out to be rather simple
because excitons are the exact one-electron-hole-pair eigenstates of the
semiconductor Hamiltonian, thus forming a complete orthogonal set for
one-pair states. It can however be of interest to extend this new
many-body theory to more complicated composite bosons, \emph{i.\ e.},
``cobosons'', which are not necessarily the one-pair eigenstates of the
system Hamiltonian, nor even orthogonal. The purpose of this paper is to
derive the ``Pauli scatterings'' and the ``interaction scatterings'' of
these cobosons formally,
\emph{i.\ e.}, just in terms of their wave functions and the interaction
potentials which exist between the fermions from which they are
constructed. We also explain how to derive many-body effects in this very
general system of composite bosons.
\end{abstract}

\newpage

A few years ago, we have tackled the difficult problem of many-body
effects between composite bosons through the study of interacting excitons
in semiconductors [1-3]. Excitons actually constitute a very nice ``toy
model'' since the semiconductor Hamiltonian is extremely simple --- just
electrons and holes with kinetic energy and Coulomb interaction
--- the full spectrum of the exciton eigenstates being
analytically known in terms of hypergeometric functions, in 3D and 2D.
When we started these studies, we had in mind to better understand the
bosonization procedures [4] and to properly determine their limit of
validity, through full-proof ab
initio calculations. To our major surprise --- and contentment --- we
have found that, whatever the effective Hamiltonians for bosonized
excitons [5] are, they miss a set of processes which actually produce the
dominant terms in various problems of major physical interest, such as the
semiconductor optical nonlinearities.

The many-body theory we have constructed, which only uses
the semiconductor Hamiltonian written in terms of free electrons and free
holes, makes appearing \emph{two} fully independent scatterings [1-3]: One
is associated to Coulomb processes between two excitons, the ``in'' and
``out'' excitons being made with the same electron-hole pairs. The second
scattering is completely novel. It directly comes from the
undistinguishability of the fermionic components of the excitons, and
describe the carrier exchanges which can take place between two excitons,
in the absence of any Coulomb process. While the direct Coulomb
scatterings $\xi$ are energy-like quantities, these novel ``Pauli
scatterings'' $\lambda$ are dimensionless, so that
they are, by construction, missed in any model Hamiltonian for
interacting excitons, \emph{whatever} the effective scatterings
of these Hamiltonians are --- a very strong statement, indeed!

From dimensional arguments only, it is possible to show [6-8] that the
semiconductor optical nonlinearities are entirely controlled by these
Pauli scatterings at large detuning, so that there is no hope to
correctly describe these nonlinearities through effective Hamiltonians
for bosonized excitons.

We could have dreamed of a better correctness in physical effects
controlled by energy-like scatterings, such as the scattering rates of
two excitons. Unfortunately, we have shown [9] that, in order to recover
the correct value of these quantities, the effective scatterings between
excitons that must be introduced in the exciton Hamiltonian, make this
Hamiltonian non hermitian --- although different from the usual exciton
Hamiltonian, --- a major physical failure hard to accept.

All this led us to the conclusion that, in order to correctly describe
many-body effects with excitons, it is not possible to ``cook'' the
Coulomb interactions between electrons and holes with carrier exchanges,
once and for all, in a set of ``Coulomb scatterings dressed by exchange''
as done in all model Hamiltonians describing interacting excitons.

Since essentially all quantum particles known as bosons are composite
particles, it can be of interest to extend our many-body theory for
excitons to any type of composite bosons, \emph{i.\ e.}, to formally
write their Pauli and interaction scatterings without using any
particular form for these composite bosons nor for the system
Hamiltonian.

The paper is organized as follows.

In the first section, we settle the notations and formally define the
arbitrary composite bosons we study in this work, through their
expansion in terms of free fermions $\alpha$ and free fermions $\beta$.

We have to consider that the cobosons form a complete set for
one-fermion-pair states in order to possibly describe any system of
fermion pairs in terms of cobosons. However, this does not impose the
one-coboson states to be orthogonal --- in connexion with one of our
recent works on electron teleportation between quantum dots [10], in which
one of the composite bosons of physical interest is a pair of trapped
electrons.

In section 2, we determine the Pauli scatterings, due to fermion
exchanges between these composite bosons. As physically reasonable, they
only depend on the composite bosons of interest, through their wave
functions, but not on the system Hamiltonian. We, in particular, show how
our results on the scalar products of exciton-states can be readily
extended to arbitrary composite bosons, even if the one-coboson states
are not orthogonal.

In section 3, we show how we can formally write the energy-like
interaction scatterings for any type of composite bosons --- not
necessarily the eigenstates of the system Hamiltonian --- in terms of the
potentials between fermions
$\alpha$ and $\beta$ appearing in this Hamiltonian.

In a last section, we explain how to derive many-body effects between
these composite bosons, following a path similar to the one we have used
for excitons.

As our works on exciton many-body effects have pointed out quite clearly
many weaknesses of the bosonization procedures, while the many-body
theory we have constructed, now allows to treat exchange
processes between composite particles exactly, it can be of interest to
introduce a new name for these tricky quantum particles, the
``coboson'' --- as a contraction of ``composite boson''.

\section{Arbitrary composite bosons}

We consider composite bosons made of one fermion $\alpha$ and one fermion
$\beta$. Let us introduce two \emph{arbitrary} orthogonal basis for
these fermions,
\begin{eqnarray}
|\v k_\alpha\rangle &=& a_{\v k_\alpha}^\dag|v\rangle\ ,\nonumber\\
|\v k_\beta\rangle &=& b_{\v k_\beta}^\dag|v\rangle\ ,
\end{eqnarray}
the anticommutators of their creation operators being such that
$\{a_{\v k'},a_{\v k}^\dag\}_+=\delta_{\v k',\v k}=\{b_{\v k'},b_{\v
k}^\dag\}_+$. 

The states $|\v k_\alpha,\v k_\beta\rangle=a_{\v
k_\alpha}^\dag  b_{\v k_\beta}^\dag|v\rangle$ form a complete set for one
fermion pair
$(\alpha,\beta)$, so that the closure relation for one-pair states 
reads
\begin{equation}
I=\sum_{\v k_\alpha,\v k_\beta}|\v k_\alpha,\v k_\beta\rangle\,
\langle\v k_\beta,\v k_\alpha|\ .
\end{equation}
This closure relation allows to write any state $|i\rangle$ made
of one $(\alpha,\beta)$ pair as
\begin{equation}
|i\rangle=\sum_{\v k_\alpha,\v k_\beta}|\v k_\alpha,\v k_\beta\rangle\,
\langle\v k_\beta,\v k_\alpha|i\rangle\ .
\end{equation}
If we now write this one-pair state $|i\rangle$ as $B_i^\dag|v\rangle$, we
readily deduce that the creation operator $B_i^\dag$ reads in
terms of creation operators for free fermions $\alpha$ and $\beta$, as
\begin{equation}
B_i^\dag=\sum_{\v k_\alpha,\v k_\beta}a_{\v k_\alpha}^\dag b_{\v
k_\beta}^\dag\,\langle\v k_\beta,\v k_\alpha|i\rangle\ .
\end{equation}
Being made of a pair of fermion operators, $B_i^\dag$ is a
composite boson creation operator.

In order to possibly describe a system of $(\alpha,\beta)$ pairs entirely
in terms of cobosons, it is necessary for these cobosons to form a
complete set for one-pair states. If the coboson states $|i\rangle$ are
normalized and orthogonal, their closure relation simply reads
\begin{equation}
I=\sum_i|i\rangle\,\langle i|\ .
\end{equation}
This allows to write the creation operator for a free fermion pair in
terms of coboson creation operators as
\begin{equation}
a_{\v k_\alpha}^\dag b_{\v k_\beta}^\dag=\sum_iB_i^\dag\,\langle i|\v
k_\alpha,\v k_\beta\rangle\ .
\end{equation}

If the cobosons of physical interest form a complete set, but if this set
is not orthogonal --- as the pairs of trapped electrons we have studied in
ref.\ [10], --- their closure relation is not as simple as eq.\ (5). It
now reads
\begin{equation}
I=\sum_{i,j}|i\rangle\,z_{ij}\,\langle j|\ ,
\end{equation}
where the prefactors $z_{ij}$ are such that
\begin{equation}
\sum_m z_{im}\,\langle m|j\rangle=\delta_{ij}\ .
\end{equation}
The above equation just says that the matrix made of the $z_{ij}$'s and
the matrix made of the $\langle i|j\rangle$'s are inverse matrices. 
For nonorthogonal cobosons, the link between free pair and coboson
creation operators is then given by
\begin{equation}
a_{\v k_\alpha}^\dag b_{\v k_\beta}^\dag=\sum_{i,j}B_i^\dag\,z_{ij}\,
\langle j|\v k_\alpha,\v k_\beta\rangle\ .
\end{equation}

\section{Coboson scattering due to fermion exchange}

The ``interactions'' between two composite bosons which only come from
the fact that these cobosons can exchange their fermions, do not depend on
the forces acting on these fermions. Consequently, to determine these
``Pauli scatterings'', it is not necessary to specify
the system Hamiltonian at hand.

\subsection{``Deviation-from-boson operator''}
By using eq.\ (4) for the coboson creation operators, we readily get from
eq.\ (2),
\begin{equation}
[B_m,B_i^\dag]=\langle m|i\rangle-D_{mi}\ ,
\end{equation}
where $D_{mi}$ is the ``deviation-from-boson operator''. This operator,
which is such that $D_{mi}|v\rangle=0$, as obtained by multiplying the
above equation by
$|v\rangle$ on the right, in fact appears as
$D_{mi}^{(\alpha)}+D_{mi}^{(\beta)}$. In the $D_{mi}^{(\alpha)}$ part,
given by
\begin{equation}
D_{mi}^{(\alpha)}=\sum_{\v k_\beta ',\v k_\beta}b_{\v k_\beta}^\dag
b_{\v k_\beta '}\sum_{\v k_\alpha}\langle m|\v k_\alpha,\v k_\beta
'\rangle\,\langle\v k_\beta,\v k_\alpha|i\rangle\ ,
\end{equation}
the cobosons $m$ and $i$ are made with the same fermion $\alpha$, while
in $D_{mi}^{(\beta)}$, given by
\begin{equation}
D_{mi}^{(\beta)}=\sum_{\v k_\alpha ',\v k_\alpha}a_{\v k_\alpha}^\dag
a_{\v k_\alpha '}\sum_{\v k_\beta}\langle m|\v k_\alpha ',\v k_\beta
\rangle\,\langle\v k_\beta,\v k_\alpha|i\rangle\ ,
\end{equation}
the cobosons $m$ and $i$ are made with the same fermion
$\beta$.

\subsection{``Pauli scatterings'' for cobosons}

To go further and deduce the ``Pauli scatterings'' between cobosons, it is
necessary to consider that these cobosons form a complete basis for
one-pair states, in order to possibly write a pair
of free fermions $(\alpha,\beta)$ in terms of cobosons, using eq.\ (6)
or (9).

\subsubsection{Orthogonal cobosons}

Let us start with orthogonal cobosons related to free pairs through eq.\
(6). The ``Pauli scatterings'' $\lambda\left(^{n\ \,j}_{m\ i}\right)$, due
to fermion exchanges between composite particles, are defined through
\begin{equation}
\left[D_{mi},B_j^\dag\right]=\sum_n\left[\lambda\left(^{n\ \,j}_{m\ i}
\right)+\lambda\left(^{m\ j}_{n\ \,i}\right)\right]\,B_n^\dag\ .
\end{equation}
To calculate $\lambda$, we use eqs.\ (4,11) to get
\begin{equation}
\left[D_{mi}^{(\alpha)},B_j^\dag\right]=\sum_{\v k_\beta ',\v k_\beta,\v
k_\alpha}\sum_{\v p_\alpha,\v p_\beta}
\langle m|\v k_\alpha,\v k_\beta '\rangle\,\langle\v k_\beta,\v
k_\alpha|i\rangle\,\langle\v p_\beta,\v p_\alpha|j\rangle\,\delta_{\v
k_\beta ',\v p_\beta}\,a_{\v p_\alpha}^\dag b_{\v k_\beta}^\dag\ .
\end{equation}
We then express $a^\dag b^\dag$ in terms of $B^\dag$ according to eq.\
(6). This leads to
\begin{equation}
\left[D_{mi}^{(\alpha)},B_j^\dag\right]=\sum_n\lambda\left(^{n\ \,j}_{m\
i}\right)\,B_n^\dag\ ,
\end{equation}
where $\lambda\left(^{n\ \,j}_{m\ i}\right)$ is given by
\begin{equation}
\lambda\left(^{n\ \,j}_{m\ i}\right)=\sum_{\v k_\alpha ',\v k_\alpha,\v
k_\beta ',\v k_\beta}\langle m|\v k_\alpha,\v k_\beta '\rangle\,
\langle n|\v k_\alpha ',\v k_\beta\rangle\,\langle\v k_\beta,\v
k_\alpha|i\rangle\,\langle\v k_\beta ',\v k_\alpha '|j\rangle\ .
\end{equation}
The second term on the RHS of eq.\ (13) is obtained in the same way, by
calculating $\left[D_{mi}^{(\beta)},B_j^\dag\right]$. 

The Pauli scattering $\lambda\left(^{n\ \,j}_{m\ i}\right)$
is shown in Fig.1a. As for composite excitons, it corresponds
to a fermion exchange between the ``in'' cobosons $(i,j)$ such that the
coboson $m$ ends by having the same fermion $\alpha$ as the coboson $i$.
(By convention, the cobosons of the lowest line of the Pauli scattering
$\lambda\left(^{n\ \,j}_{m\ i}\right)$, here $m$ and $i$, are made with
the same fermion $\alpha$).

We can rewrite this Pauli scattering in $\v r$ space by using
\begin{equation}
\langle\v k_\beta,\v k_\alpha|i\rangle=\int d\v r_\alpha\,d\v r_\beta\,
\langle\v k_\beta|\v r_\beta\rangle\,\langle\v k_\alpha|\v
r_\alpha\rangle\,\langle\v r_\beta,\v r_\alpha|i\rangle\ ,
\end{equation}
and by performing the summation over the various $\v k$'s through closure
relations. We find that $\lambda\left(^{n\ \,j}_{m\ i}\right)$ 
reads in terms of the wave functions of the $(m,n)$ and $(i,j)$ cobosons,
as
\begin{equation}
\lambda\left(^{n\ \,j}_{m\ i}\right)=\int d\v r_{\alpha_1}\,d\v
r_{\alpha_2}\,d\v r_{\beta_1}\,d\v r_{\beta_2}\,
\phi_m^\ast(\v r_{\alpha_1},\v r_{\beta_2})\,\phi_n^\ast(\v r_{\alpha_2},
\v r_{\beta_1})\,\phi_i(\v r_{\alpha_1},\v r_{\beta_1})\,
\phi_j(\v r_{\alpha_2},\v r_{\beta_2})\ ,
\end{equation}
where $\phi_i(\v r_{\alpha},\v r_{\beta})=\langle\v r_\beta,\v r_\alpha
|i\rangle$ is the wave function of the coboson $i$ (see Fig.2a). 

\subsubsection{Nonorthogonal cobosons}

If the cobosons form a nonorthogonal basis for one-pair states, the link
between the creation operators for free fermion pairs and cobosons given
in eq.\ (6) has to be replaced by the link given in eq.\ (9). From it, we
now get
\begin{equation}
\left[D_{mi},B_j^\dag\right]=\sum_nB_n^\dag\sum_p\,z_{np}\,\left[
\lambda\left(^{p\ \,j}_{m\ i}\right)+\lambda\left(^{m\ j}_{p\ \,i}
\right)\right]\ ,
\end{equation}
where $\lambda\left(^{p\ \,j}_{m\ i}\right)$ is the same Pauli scattering
as the one defined in eq.\ (16) or (18).

\subsection{Scalar product of coboson states}

Equation (19) for cobosons forming a nonorthogonal basis is definitely not
as simple as eq.\ (13). This, however, has no major consequence on the
scalar product of
$N$-coboson states. Indeed, if we consider the scalar product of two
coboson states, we find, using eq.\ (10),
\begin{equation}
\langle v|B_mB_nB_i^\dag B_j^\dag|v\rangle=\langle n|i\rangle\,\langle
m|j\rangle+\langle n|j\rangle\,\langle m|i\rangle-
\langle v|B_mD_{ni}B_j^\dag|v\rangle\ ,
\end{equation}
the last term of the above equation reading, due to eq.\ (19),
\begin{equation}
\langle v|B_mD_{ni}B_j^\dag|v\rangle=\sum_{p,q}\langle m|p\rangle\,
z_{pq}\,\left[\lambda\left(^{q\ j}_{n\ i}\right)+\lambda\left(^{n\ j}
_{q\ i}\right)\right]\ .
\end{equation}
So that, due to eq.\ (8), the scalar product of two-coboson states reduces
to
\begin{equation}
\langle v|B_mB_nB_i^\dag B_j^\dag|v\rangle=\langle n|i\rangle\,\langle
m|j\rangle+\langle n|j\rangle\,\langle m|i\rangle-
\lambda\left(^{n\ \,j}_{m\ i}\right)-\lambda\left(^{m\ j}_{n\ \,i}
\right)\ .
\end{equation}
The exchange part of this scalar product is just the one for
orthogonal cobosons --- or for excitons [1] ---, the only difference
coming from the na\"{\i}ve part, \emph{i.\ e.}, the part which remains
when cobosons are replaced by elementary particles, the scalar product
$\langle m|i\rangle$ being just replaced by $\delta_{m,i}$ if the
cobosons are orthogonal.

It is possible to show that this nicely simple result can be extended to
more complicated scalar products of coboson states.

\section{Coboson scatterings due to interactions between fermions}

The cobosons interact through fermion exchanges as described in the
preceding section. They also interact, in a more standard way, through the
forces which exist between the fermions from which they are constructed.
It is of importance to note that this second coboson interaction, which
can appear as rather na\"{\i}ve, is in fact very subtle due to the fermion
undistinguishability. Indeed, with fermions
$(\alpha_1,\alpha_2,\beta_1,\beta_2)$, two kinds of cobosons can be made,
$(\alpha_1,\beta_1)$ and $(\alpha_2,\beta_2)$, or $(\alpha_1,\beta_2)$
and $(\alpha_2,\beta_1)$. Due to this, the interactions \emph{between}
cobosons resulting from forces between fermions $\alpha$ and $\beta$,
must be taken as $v(\alpha_1,\beta_2)+v(\alpha_2,\beta_1)$ in the first
case, but $v(\alpha_1,\beta_1)+v(\alpha_2,\beta_2)$ in the second case.
Since there is no way to know with which pairs of fermions the cobosons
are made, there is no way to unambiguously write the interactions
\emph{between} cobosons associated to the forces between fermions
$\alpha$ and $\beta$.

It is however clear that, even if the interactions between cobosons due
to forces between fermions $\alpha$ and $\beta$ cannot be properly
defined, these forces must play a role in the many-body physics of these
cobosons. The clean way to make them appearing is actually non standard.
It again relies on a set of commutators.

\subsection{System Hamiltonian for fermions $\alpha$ and $\beta$}

The general form for a system Hamiltonian made of
fermions
$\alpha$ and
$\beta$ reads in first quantization as
\begin{equation}
H=H_\alpha+H_\beta+V_{\alpha\alpha}+V_{\beta\beta}+V_{\alpha\beta}\ .
\end{equation}
$H_\alpha$ and $H_\beta$ are one-body operators for fermions $\alpha$ and
fermions $\beta$:
\begin{equation}
H_\alpha=\sum_nh_\alpha(\v r_{\alpha n})\hspace{2cm} H_\beta=\sum_n
h_\beta(\v r_{\beta n})\ .
\end{equation}
The three other terms of the Hamiltonian (23) are two-body operators
which correspond to interactions between cobosons $\alpha$, between
cobosons $\beta$ and between cobosons $\alpha$ and $\beta$:
\begin{equation}
V_{\alpha\alpha}=\frac{1}{2}\sum_{n\neq n'}v_{\alpha\alpha}(\v
r_{\alpha n},\v r_{\alpha n'})\ ,
\end{equation}
\begin{equation}
V_{\alpha\beta}=\sum_{n,n'}v_{\alpha\beta}(\v r_{\alpha n},\v
r_{\beta n'})\ .
\end{equation}

In terms of the creation operators for the free fermion states introduced
in section 2 (which, in general, are not the exact eigenstates of
$h_\alpha$ and $h_\beta$), the non-interacting part of the system
Hamiltonian reads
\begin{equation}
H_\alpha=\sum_{\v k_\alpha,\v k_\alpha '}\langle\v k_\alpha '|h_\alpha|
\v k_\alpha\rangle\,a_{\v k_\alpha '}^\dag a_{\v k_\alpha}\ ,
\end{equation}
\begin{equation}
H_\beta=\sum_{\v k_\beta,\v k_\beta '}\langle\v k_\beta '|h_\beta|
\v k_\beta\rangle\,b_{\v k_\beta '}^\dag b_{\v k_\beta}\ ,
\end{equation}
where the prefactors are given by
\begin{equation}
\langle\v k_\alpha '|h_\alpha|\v k_\alpha\rangle=\int d\v r\,\langle\v
k_{\alpha}'|\v r\rangle\,h_\alpha(\v r)\,\langle\v r|\v k_\alpha\rangle\ .
\end{equation}
and similarly for $\langle\v k_\beta '|h_\beta|\v k_\beta\rangle$.
In the same way, the two-body interacting parts of the Hamiltonian $H$
read in second quantization, on this basis, as
\begin{equation}
V_{\alpha\alpha}=\frac{1}{2}\sum_{\v k_{\alpha}',\v k_{\alpha},
\v q_{\alpha}',\v q_{\alpha}}v_{\alpha\alpha}\left(^{\v q_{\alpha}'\ \v
q_{\alpha}}_{\v k_{\alpha}'\ \v k_{\alpha}}\right)\, a_{\v
k_{\alpha}'}^\dag a_{\v q_{\alpha}'}^\dag a_{\v q_{\alpha}} a_{\v
k_{\alpha}}\ ,
\end{equation}
\begin{equation}
V_{\alpha\beta}=\sum_{\v k_{\alpha}',\v k_{\alpha},
\v k_{\beta}',\v k_{\beta}}v_{\alpha\beta}\left(^{\v k_{\beta}'\ \v
k_{\beta}}_{\v k_{\alpha}'\ \v k_{\alpha}}\right)\, a_{\v
k_{\alpha}'}^\dag b_{\v k_{\beta}'}^\dag b_{\v k_{\beta}} a_{\v
k_{\alpha}}\ ,
\end{equation}
where the prefactors are given by
\begin{eqnarray}
v_{\alpha\alpha}\left(^{\v q_{\alpha}'\ \v q_{\alpha}}_{\v k_{\alpha}'\
\v k_{\alpha}}\right) &=&
\int d\v r_{\alpha}\,d\v r_{\alpha}'\,\langle\v k_{\alpha}'|\v
r_{\alpha}\rangle\,\langle\v q_{\alpha}'|\v r_{\alpha}'\rangle\,
v_{\alpha\alpha}(\v r_{\alpha},\v r_{\alpha}')\,
\langle\v r_{\alpha}'|\v q_{\alpha}\rangle\,
\langle\v r_{\alpha}|\v k_{\alpha}\rangle\nonumber\\
&=& v_{\alpha\alpha}\left(_{\v q_{\alpha}'\ \v q_{\alpha}}^{\v
k_{\alpha}'\
\v k_{\alpha}}\right)\ ,
\end{eqnarray}
and similarly for the other prefactors.

\subsection{Orthogonal cobosons}

\subsubsection{``Creation potential''}

Let us first consider cobosons forming an orthogonal set, these cobosons
being not necessarily the exact one-pair eigenstates of the system
Hamiltonian. Due to the closure relation (5) for orthogonal
states, $H$ acting on $|i\rangle$ then reads
\begin{equation}
H|i\rangle=\sum_m|m\rangle\,\langle m|H|i\rangle\ ,
\end{equation}
with $\langle m|H|i\rangle=E_i\,\delta_{m,i}$ if the
cobosons are eigenstates of the system Hamiltonian,
\emph{i.\ e.}, if $(H-E_i)|i\rangle=0$. Equation (33) leads to define the
``creation potential''
$V_i^\dag$ for the coboson $i$ as
\begin{equation}
[H,B_i^\dag]=\sum_m\langle m|H|i\rangle\,B_m^\dag+V_i^\dag\ ,
\end{equation}
in order for the creation potential to be such that
\begin{equation}
V_i^\dag|v\rangle=0\ .
\end{equation}
This insures $V_i^\dag$ to indeed describe the interactions of the
coboson $i$ with the rest of the system. Let us now calculate
this $V_i^\dag$ explicitly.

\noindent (i) It is possible to
split the commutator of $B_i^\dag$, given in eq.\ (4), with the part of
the Hamiltonian acting on fermion pairs, into three terms:
\begin{equation}
\left[H_\alpha+H_\beta+V_{\alpha\beta},B_i^\dag\right]=A_1^\dag
+A_2^\dag+A_3^\dag\ ,
\end{equation}
with $A_1^\dag$ in $a^\dag b^\dag$, $A_2^\dag$ in $a^\dag b^\dag b^\dag
b$ and $A_3^\dag$ in $a^\dag b^\dag a^\dag a$. 
The first term $A_1^\dag$ precisely reads
\begin{eqnarray}
A_1^\dag = \sum_{\v k_\alpha ',\v p_\beta}a_{\v k_\alpha '}^\dag
b_{\v p_\beta}^\dag\,\sum_{\v p_\alpha}\langle\v k_\alpha '|h_\alpha|
\v p_\alpha\rangle\,\langle\v p_\beta,\v p_\alpha|i\rangle
+\sum_{\v k_\beta ',\v p_\alpha}a_{\v p_\alpha}^\dag
b_{\v k_\beta '}^\dag\,\sum_{\v p_\beta}\langle\v k_\beta '|h_\beta|
\v p_\beta\rangle\,\langle\v p_\beta,\v p_\alpha|i\rangle
\nonumber\\ +\sum_{\v k_\alpha ',\v k_\beta '}a_{\v k_\alpha '}^\dag
b_{\v k_\beta '}^\dag\,\sum_{\v p_\alpha,\v p_\beta}v_{\alpha\beta}
\left(^{\v k_\beta '\ \v p_\beta}_{\v k_\alpha '\ \v p_\alpha}\right)
\,\langle\v p_\beta,\v p_\alpha|i\rangle\ .
\end{eqnarray}
By noting that $H_\alpha$ does not act on fermion $\beta$, while the
prefactor of the last term is nothing but
$$ v_{\alpha\beta}
\left(^{\v k_\beta '\ \v p_\beta}_{\v k_\alpha '\ \v
p_\alpha}\right)=
\langle\v k_\beta ',\v k_\alpha '|V_{\alpha\beta}|
\v p_\alpha,\v p_\beta\rangle\ ,$$
it is easy to see that $A_1^\dag$ can be rewritten in a compact form as
\begin{equation}
A_1^\dag=\sum_{\v k_\alpha ',\v k_\beta '} a_{\v k_\alpha '}^\dag b_{\v
k_\beta '}^\dag\,\sum_{\v p_\alpha,\v p_\beta}Ê\langle\v k_\beta ',\v
k_\alpha '|H_\alpha+H_\beta+V_{\alpha\beta}|\v p_\alpha,\v
p_\beta\rangle\,\langle\v p_\beta,\v p_\alpha|i\rangle\ .
\end{equation}
If we now use eq.\ (6) to write $a^\dag b^\dag$ in terms of cobosons,
we end, due to eq.\ (2), with
\begin{equation}
A_1^\dag=\sum_m\langle m|H|i\rangle\, B_m^\dag\ ,
\end{equation}
which is just the first term of eq.\ (34).

The second term on the RHS of eq.\ (36), $A_2^\dag$, appears as
\begin{equation}
A_2^\dag=\sum_{\v k_\alpha ',\v p_\beta} a_{\v k_\alpha '}^\dag
b_{\v p_\beta}^\dag\,\sum_{\v k_\beta ',\v k_\beta}b_{\v k_\beta '}^\dag
b_{\v k_\beta}\,\sum_{\v p_\alpha}v_{\alpha\beta}\left(
^{\v k_\beta '\ \v k_\beta}_{\v k_\alpha '\ \v p_\alpha}\right)\,\langle
\v p_\beta,\v p_\alpha|i\rangle\ .
\end{equation}
We rewrite the first $a^\dag b^\dag$ in terms of coboson operators
according to eq.\ (6), to make $A_2^\dag$ reading as
$$A_2^\dag=\sum_mB_m^\dag\,\sum_{\v k_\beta ',\v k_\beta}b_{\v k_\beta '}
^\dag b_{\v k_\beta}\,X_{\alpha\beta}(\v k_\beta ',\v k_\beta;m,i)\ ,$$
\begin{equation}
X_{\alpha\beta}(\v k_\beta ',\v k_\beta;m,i)=\sum_{\v k_\alpha ',\v
p_\alpha,\v p_\beta}\langle m|\v k_\alpha ',\v p_\beta\rangle\,
v_{\alpha\beta}\left(
^{\v k_\beta '\ \v k_\beta}_{\v k_\alpha '\ \v p_\alpha}\right)\,\langle
\v p_\beta,\v p_\alpha|i\rangle\ .
\end{equation}
In the same way, $A_3^\dag$ is found to be
$$A_3^\dag=\sum_mB_m^\dag\,\sum_{\v k_\alpha ',\v k_\alpha}a_{\v k_\alpha
'} ^\dag a_{\v k_\alpha}\,Y_{\alpha\beta}(\v k_\alpha ',\v k_\alpha;m,i)\
,$$
\begin{equation}
Y_{\alpha\beta}(\v k_\alpha ',\v k_\alpha;m,i)=\sum_{\v k_\beta ',\v
p_\alpha,\v p_\beta}\langle m|\v p_\alpha,\v k_\beta '\rangle\,
v_{\alpha\beta}\left(
^{\v k_\beta '\ \v p_\beta}_{\v k_\alpha '\ \v k_\alpha}\right)\,\langle
\v p_\beta,\v p_\alpha|i\rangle\ .
\end{equation}

\noindent (ii) If we now turn to the interactions between fermions
$\alpha$, the same procedure leads to
$$\left[V_{\alpha\alpha},B_i^\dag\right]=
\sum_mB_m^\dag\,\sum_{\v k_{\alpha}',\v k_{\alpha}}a_{\v k_{\alpha}
'} ^\dag a_{\v k_{\alpha}}\,Y_{\alpha\alpha}(\v k_{\alpha}',\v
k_{\alpha};m,i)\ ,$$
\begin{equation}
Y_{\alpha\alpha}(\v k_{\alpha}',\v k_{\alpha};m,i)=\sum_{\v
q_{\alpha}',\v p_\alpha,\v p_\beta}\langle m|\v q_{\alpha}',\v
p_\beta\rangle\, v_{\alpha\alpha}\left( _{\v q_{\alpha}'\ \v
p_\alpha}^{\v k_{\alpha}'\ \v k_{\alpha}}\right)\,\langle
\v p_\beta,\v p_\alpha|i\rangle\ ,
\end{equation}
while the interactions between fermions $\beta$ give
$$\left[V_{\beta\beta},B_i^\dag\right]=
\sum_mB_m^\dag\,\sum_{\v k_{\beta} ',\v k_{\beta}}b_{\v k_{\beta}
'} ^\dag b_{\v k_{\beta}}\,X_{\beta\beta}(\v k_{\beta}',\v
k_{\beta};m,i)\ ,$$
\begin{equation}
X_{\beta\beta}(\v k_{\beta}',\v k_{\beta};m,i)=\sum_{\v q_{\beta}',\v
p_\alpha,\v p_\beta}\langle m|\v p_{\alpha},\v q_{\beta}'\rangle\,
v_{\beta\beta}\left( _{\v q_{\beta}'\ \v p_\beta}^{\v k_{\beta}'\ \v
k_{\beta}}\right)\,\langle
\v p_\beta,\v p_\alpha|i\rangle\ ,
\end{equation}

\noindent (iii) By collecting the results of eqs.\ (36,39,41,42-44), the
creation potential
$V_i^\dag$, defined in eq.\ (34), finally reads
$$V_i^\dag=\sum_mB_m^\dag\,W_{mi}\ ,$$
where the operator $W_{mi}$ is defined by
\begin{eqnarray}
W_{mi} &=& \sum_{\v k_\alpha',\v k_\alpha}
a_{\v k_\alpha'}^\dag a_{\v
k_\alpha}\left[Y_{\alpha\alpha}(\v k_\alpha',\v
k_\alpha;m,i)+Y_{\alpha\beta}(\v k_\alpha',\v k_\alpha;m,i)
\right]\nonumber\\  
&+& \sum_{\v k_\beta',\v k_\beta}
b_{\v k_\beta'}^\dag b_{\v k_\beta}\left[X_{\beta\beta}(\v
k_\beta',\v k_\beta;m,i)+X_{\alpha\beta}(\v k_\beta',\v
k_\beta;m,i)\right]\ .
\end{eqnarray}
Since $W_{mi}|v\rangle=0$, it is thus easy to check that the condition
(35) for a creation potential, is indeed fulfilled by this $V_i^\dag$.

\subsubsection{``Interaction scatterings''}

The ``direct interaction scatterings'' between cobosons $i$ and $j$
physically come from the interactions between fermions $(\alpha,\alpha)$,
between fermions $(\beta,\beta)$ and also from the interactions between
fermions
$(\alpha,\beta)$, with the part between the fermions $(\alpha,\beta)$
making the coboson $i$ excluded. These interaction scatterings are
formally defined through 
\begin{equation}
\left[V_i^\dag,B_j^\dag\right]=\sum_{mn}\xi\left(^{n\ \,j}_{m\ i}\right)\,
B_m^\dag B_n^\dag\ ,
\end{equation}
so that these scatterings are such that
\begin{equation}
\left[W_{mi},B_j^\dag\right]=\sum_n \xi\left(^{n\ \,j}_{m\ i}\right)\,
B_n^\dag\ .
\end{equation}
To calculate $\xi$, let us consider the first term of eq.\ (45), in
$Y_{\alpha\alpha}$. Using eqs.\ (4,6), the commutator of
this first term with $B_j^\dag$ reads
\begin{eqnarray}
[W_{mi}^{(1)},B_j^\dag] &=& \sum_{\v k_\alpha ',\v k_\alpha}\sum_{\v
p_{\alpha}',\v p_{\beta}'}Y_{\alpha\alpha}(\v k_\alpha ',\v
k_\alpha;m,i)\,\langle\v p_{\beta}',\v p_{\alpha}'|j\rangle\,
[a_{\v k_\alpha '}^\dag a_{\v k_\alpha},a_{\v p_{\alpha}'}^\dag b_{\v
p_{\beta}'}^\dag]\nonumber\\ &=&
\sum_{\v k_\alpha ',\v p_\alpha ',\v p_{\beta}'}
Y_{\alpha\alpha}(\v k_\alpha ',\v
p_\alpha ';m,i)\,\langle\v p_{\beta}',\v p_{\alpha}'|j\rangle\,
\sum_nB_n^\dag\,\langle n|\v k_\alpha ',\v p_\beta '\rangle\ .
\end{eqnarray}
By inserting $Y_{\alpha\alpha}$ given in eq.\ (43) into the above
equation, we find that the first term of $\xi\left(^{n\ \,j}_{m\
i}\right)$ reads
\begin{equation}
\xi_1\left(^{n\ \,j}_{m\ i}\right)=
\sum_{\v k_\alpha ',\v q_\alpha ',\v p_{\alpha}',
\v p_{\beta}',\v p_\alpha,\v p_\beta}
\langle m|\v q_{\alpha}',\v p_\beta\rangle\,
\langle n|\v k_\alpha ',\v p_{\beta}'\rangle\,
v_{\alpha\alpha}\left(_{\v q_{\alpha}'\ \v p_\alpha}^{\v k_{\alpha}'\
\v p_{\alpha}'}\right)\,
\langle\v p_\beta,\v p_\alpha|i\rangle\,
\langle\v p_{\beta}',\v p_{\alpha}'|j\rangle\ .
\end{equation}
This $\xi_1\left(^{n\ \,j}_{m\ i}\right)$ is shown in Fig.1b. It
corresponds to an interaction between the fermions $\alpha$ of the ``in''
cobosons $(i,j)$, the ``out'' cobosons $(m,n)$ being made with the same
fermion pair as the ``in'' cobosons. We can rewrite this $\xi_1$ in real
space, by using eqs.\ (17) and (32) and by performing the
summation over the various
$\v k$'s through closure relations. This leads to
\begin{equation}
\xi_1\left(^{n\ \,j}_{m\ i}\right)=\int d\v r_{\alpha 1}d\v r_{\alpha
2} d\v r_{\beta 1}d\v r_{\beta 2}\phi_m^\ast(\v r_{\alpha 1},\v
r_{\beta 1})\phi_n^\ast(\v r_{\alpha 2},\v r_{\beta 2})
v_{\alpha\alpha}(\v r_{\alpha 1},\v r_{\alpha 2})
\phi_i(\v r_{\alpha 1},
\v r_{\beta 1})\phi_j(\v r_{\alpha 2},\v r_{\beta 2})\ ,
\end{equation}
which is shown in Fig.1c.

By calculating the contributions of
the three other terms of eq.\ (45) in the same way, we end with a direct
interaction scattering which has a form very
similar to the one for excitons [1], namely
\begin{eqnarray}
\xi\left(^{n\ \,j}_{m\ i}\right)=\int d\v r_{\alpha 1}\,d\v r_{\alpha 2}\,
d\v r_{\beta 1}\,d\v r_{\beta 2}\,\phi_m^\ast(\v r_{\alpha 1},\v r_{\beta
1})\,\phi_n^\ast(\v r_{\alpha 2},\v r_{\beta 2})\,\phi_i(\v r_{\alpha 1},
\v r_{\beta 1})\,\phi_j(\v r_{\alpha 2},\v r_{\beta 2})\nonumber\\
\times\ \left[
v_{\alpha\alpha}(\v r_{\alpha 1},\v r_{\alpha 2})+
v_{\beta\beta}(\v r_{\beta 1},\v r_{\beta 2})+
v_{\alpha\beta}(\v r_{\alpha 1},\v r_{\beta 2})+
v_{\alpha\beta}(\v r_{\alpha 2},\v r_{\beta 1})\right]\ .
\end{eqnarray}
This direct scattering is represented in Fig.2b: In it, no fermion
exchange takes place between the ``in'' cobosons $(i,j)$.

\subsubsection{Matrix elements of the system Hamiltonian in the 2-coboson
subspace}

Using the commutators given in eqs.\ (34,46), the Hamiltonian in the
two-coboson subspace appears as
\begin{eqnarray}
\langle v|B_mB_n\,H\,B_i^\dag B_j^\dag|v\rangle=\sum_q
\langle v|B_mB_nB_i^\dag B_q^\dag|v\rangle\,\langle q|H|j\rangle\ +\
(i\leftrightarrow j)\nonumber\\
+\sum_{pq}\xi\left(^{q\ j}_{p\ i}\right)\,\langle v|B_mB_nB_p^\dag
B_q^\dag|v\rangle\ .
\end{eqnarray}
So that, using eq.\ (22 ) for the scalar product of orthogonal cobosons,
we end with
\begin{eqnarray}
\langle v|B_mB_n\,H\,B_i^\dag
B_j^\dag|v\rangle=\left[\delta_{m,i}\,\langle n|H|j\rangle+\delta_{n,i}\,
\langle m|H|j\rangle-\sum_q\left(\lambda\left(^{n\ \,q}_{m\ i}\right)
+\lambda\left(^{m\ q}_{n\ \,i}\right)\right)\langle q|H|j\rangle\right.
\nonumber\\
\left.+\xi\left(^{n\ \,j}_{m\ i}\right)
-\xi^\mathrm{in}\left(^{n\ \,j}_{m\ i}\right)\right]\ +[i\leftrightarrow
j]\ ,
\end{eqnarray}
where $\xi^\mathrm{in}$ is the exchange interaction scattering shown in
Fig.2c. It is precisely given by
\begin{eqnarray}
\xi^\mathrm{in}\left(^{n\ \,j}_{m\ i}\right)&=&\sum_{pq}\lambda\left(
^{n\ \,q}_{m\ p}\right)\,\xi\left(^{q\ j}_{p\ i}\right)\nonumber\\
&=&\int d\v r_{\alpha 1}\,d\v r_{\beta 1}\,d\v r_{\alpha 2}\,d\v r_{\beta
2}\,\langle m|\v r_{\alpha 1},\v r_{\beta 2}\rangle\,\langle n|\v
r_{\alpha 2},\v r_{\beta 1}\rangle\,\langle \v r_{\beta 1},\v
r_{\alpha 1}|i\rangle\,\langle\v r_{\beta 2},\v r_{\alpha
2}|j\rangle\nonumber\\
&\times& \ \left[v_{\alpha\alpha}(\v r_{\alpha 1},\v r_{\alpha 2})
+v_{\beta\beta}(\v r_{\beta 1},\v
r_{\beta 2})+v_{\alpha\beta}(\v r_{\alpha 1},\v r_{\beta
2})+v_{\alpha\beta}(\v r_{\alpha 2},
\v r_{\beta 1})\right]\ .
\end{eqnarray}
In the case of coboson eigenstates of the system Hamiltonian, $\langle
n|H|j\rangle=E_j\,\delta_{n,j}$, we readily recover the result for
composite excitons [1].

\subsection{Nonorthogonal cobosons}

If the coboson states form a complete set for one-fermion-pair states,
but if these states are not orthogonal, we expect the preceding results
to be far more complicated. It turns out that, as for the scalar product
of cobosons, given in eq.\ (22), the only difference with the results for
orthogonal cobosons comes from the na\"{\i}ve part.

\subsubsection{Creation potential and interaction scattering}

Let us briefly go through the same path as the one we used in the
preceding subsection, the closure relation for cobosons being now given
by eq.\ (7).

The definition of the creation potential for the coboson $i$, given in
eq.\ (34) for orthogonal cobosons, now reads
\begin{equation}
\left[H,B_i^\dag\right]=\sum_mB_m^\dag\,\sum_qz_{mq}\,\langle
q|H|i\rangle+V_i^\dag\ ,
\end{equation}
in order to still have $V_i^\dag|v\rangle=0$. From the precise
calculation of $V_i^\dag$, we can then show, by using eq.\ (9) to write
$a^\dag b^\dag$ in terms of cobosons, that
\begin{equation}
V_i^\dag=\sum_{mq}B_m^\dag\,z_{mq}\,W_{qi}\ , 
\end{equation}
\begin{equation}
\left[V_i^\dag,B_j^\dag\right]=\sum_{mn}B_m^\dag B_n^\dag\sum_{pq}
z_{mp}\,z_{nq}\,\xi\left(^{q\ j}_{p\ i}\right)\ ,
\end{equation}
where $W_{qi}$ is defined in eq.\ (45) and $\xi$ is the direct interaction
scattering defined in eq.\ (51).

\subsubsection{Matrix elements of the system Hamiltonian in the 2-coboson
subspace}

Using eqs.\ (55,57), we readily find
\begin{equation}
\langle v|B_mB_nHB_i^\dag B_j^\dag|v\rangle=\langle v|B_mB_nB_i^\dag H
B_j^\dag|v\rangle\ +(i\leftrightarrow j)\ +\sum_{pqtu}
\langle v|B_mB_nB_p^\dag B_q^\dag|v\rangle\,z_{pt}z_{qu}\,\xi\left(
^{u\ j}_{t\ i}\right)\ .
\end{equation}
The closure relation (7) allows to write the first term of the above
equation as 
\begin{equation}
\langle v|B_mB_nB_i^\dag HB_j^\dag|v\rangle=\sum_{pq}\langle
v|B_mB_nB_i^\dag B_p^\dag|v\rangle\,z_{pq}\,\langle q|H|j\rangle\ ,
\end{equation}
so that, using the scalar product of coboson states given in eq.\ (22),
this first term reads
\begin{equation}
\langle v|B_mB_nB_i^\dag HB_j^\dag|v\rangle=\langle n|i\rangle\,\langle
m|H|j\rangle+\langle m|i\rangle\,\langle n|H|j\rangle-\sum_{pq}
\left[\lambda\left(^{n\ \,p}_{m\ i}\right)+\lambda\left(^{m\ p}_{n\ \,i}
\right)\right]z_{pq}\,\langle q|H|j\rangle\ ,
\end{equation}
If we now turn to the last term of eq.\ (58), we find, from the same eq.\
(22),
\begin{eqnarray}
\sum_{pqtu}
\langle v|B_mB_nB_p^\dag B_q^\dag|v\rangle\,z_{pt}z_{qu}\,\xi\left(
^{u\ j}_{t\ i}\right)=\sum_{pqtu}\langle m|p\rangle\,\langle n|q\rangle\,
z_{pt}\,z_{qu}\,\xi\left(^{u\ j}_{t\ i}\right)\hspace{3cm}\nonumber\\
-\sum_{pqtu}\lambda\left(^{n\ \,q}_{m\ p}\right)\,z_{pt}\,z_{qu}\,
\xi\left(^{u\ j}_{t\ i}\right)\ +\ (m\leftrightarrow n)\ .
\end{eqnarray}
The first term readily gives $\xi\left(^{n\ \,j}_{m\ i}\right)$, due to
eq.\ (8). By using the expressions of $\lambda$ and $\xi$ in $\v r$ space
given in eqs.\ (18,51), the second term appears as
\begin{eqnarray}
\sum_{pqtu}\lambda\left(^{n\ \,q}_{m\ p}\right)\,z_{pt}\,z_{qu}\,
\xi\left(^{u\ j}_{t\ i}\right)=\sum_{pqtu}z_{pt}\,z_{qu}\int \{d\v r\}\,
d\{\v r'\}\,\langle m|\v r_{\alpha 1}',\v r_{\beta_2}'\rangle\,\langle n|
\v r_{\alpha 2}',\v r_{\beta 1}'\rangle\, \langle\v r_{\beta 1}'\v
r_{\alpha 1}'|p\rangle\nonumber\\
\times\,
\langle\v r_{\beta 2}',\v r_{\alpha
2}'|q\rangle\,\langle t|\v r_{\alpha 1},\v r_{\beta 1}\rangle\,
\langle u|\v r_{\alpha 2},\v r_{\beta 2}\rangle\left[
v_{\alpha\alpha}(\v r_{\alpha 1},\v r_{\alpha 2})+v_{\beta\beta}(\v
r_{\beta 1},\v r_{\beta 2})\right.\nonumber\\
\left.+v_{\alpha\beta}
(\v r_{\alpha 1},\v r_{\beta 2})+v_{\alpha\beta}(\v r_{\alpha 2},\v
r_{\beta 1})\right]\,\langle\v r_{\beta 1},\v r_{\alpha 1}|i\rangle\,
\langle\v r_{\beta 2},\v r_{\alpha 2}|j\rangle\ .\hspace{1cm}
\end{eqnarray}
The summations over $(p,t)$ and $(q,u)$ being performed through eq.\ (7),
we readily find that the sum (62) reduces to the exchange interaction
scattering $\xi^\mathrm{in}\left(^{n\ \,j}_{m\ i}\right)$ given in eq.\
(54).

The above results thus show that the matrix elements of the system
Hamiltonian in an arbitrary two-coboson subspace are given by
\begin{eqnarray}
\langle v|B_mB_nHB_i^\dag B_j^\dag|v\rangle=\left\{\left[\langle
n|i\rangle\,\langle m|H|j\rangle-\sum_{pq}\lambda\left(^{n\ \,p}_{m\
i}\right)\,z_{pq}\,\langle q|H|j\rangle\right]+[i\leftrightarrow
j]\right.\nonumber\\
\left.+\xi\left(^{n\ \,j}_{m\ i}\right)-\xi^\mathrm{in}\left(^{n\ \,j}_{m\
i}\right)\right\}+\{m\leftrightarrow n\}\ .
\end{eqnarray}
This result, which reduces to eq.\ (53) when the coboson states are
orthogonal, again show that the part coming from interactions between
cobosons is formally the same whatever is the complete set of states these
cobosons form.

\section{Many-body effects with arbitrary cobosons}

The standard way to derive many-body effects between elementary quantum
particles for which the system Hamiltonian splits as $H=H_0+V$,
goes through the iteration of
\begin{equation}
\frac{1}{a-H}=\frac{1}{a-H_0}+\frac{1}{a-H}\,V\,\frac{1}{a-H_0}\ .
\end{equation}
We have shown [2] that, in the case of composite excitons which are
eigenstates of the semiconductor Hamiltonian, the equivalent of
$H=H_0+V$, deduced from $[H,B_i^\dag]=E_iB_i^\dag+V_i^\dag$, is
$HB_i^\dag=(H+E_i)B_i^\dag+V_i^\dag$, so that the equivalent of eq.\ (64)
reads
\begin{equation}
\frac{1}{a-H}\,B_i^\dag=B_i^\dag\,\frac{1}{a-H-E_i}+\frac{1}{a-H}\,V_i^\dag
\,\frac{1}{a-H-E_i}\ .
\end{equation}

In the most general case considered in this work, the creation potential
of the coboson $i$ is defined through
\begin{eqnarray}
[H,B_i^\dag]&=&\sum_m B_m^\dag\sum_q z_{mq}\,\langle
q|H|i\rangle+V_i^\dag\nonumber\\ 
&=& H_{ii}\,B_i^\dag+v_i^\dag+V_i^\dag\ ,
\end{eqnarray}
where we have set
$$H_{mi}=\sum_q z_{mq}\,\langle q|H|i\rangle\ \ \mathrm{and}\ \ v_i^\dag
=\sum_{m\neq i} B_m^\dag\,H_{mi}\ .$$
$v_i^\dag$ is such that $v_i^\dag|v\rangle\neq 0$, while
$[v_i^\dag,B_j ^\dag]=0$. The equivalent of eq.\ (65) then reads
\begin{equation}
\frac{1}{a-H}\,B_i^\dag=B_i^\dag\frac{1}{a-H-H_{ii}}+\frac{1}{a-H}\,
(v_i^\dag+V_i^\dag)\,\frac{1}{a-H-H_{ii}}\ .
\end{equation}

Using eq.\ (67) which is not far more complicated than eq.\ (65), we can
follow the same procedure as the one used for excitons, to deduce the
part of the many-body effects between arbitrary cobosons coming from
interactions \emph{between} the elementary fermions making these composite
bosons. Their correlations read in terms of matrix elements between
$N$-coboson states which look like
$$\langle v|B_{m_N}\cdots B_{m_1}\,\frac{1}{a-H}\,B_{i_1}^\dag\cdots
B_{i_N}^\dag|v\rangle\ .$$
To calculate them, we first push $1/(a-H)$ to the right according to
eq.\ (67) and we eliminate the various ``creation potentials'' through
eq.\ (46) or (57). This makes appearing a lot of interaction scatterings
$\xi$. We end with scalar products of
$N$-coboson states, which do not contain the system Hamiltonian anymore.
These scalar products are then calculated, as for excitons, in terms of
Pauli scatterings between two cobosons, using eq.\ (10) and eq.\ (13) or
(19) --- as done to get eq.\ (22) for just $N=2$ cobosons. When $N$ is
large, these scalar products are better represented by Shiva diagrams for
fermion exchanges between $P$ excitons, with $2\leq P\leq N$, as
explained more in details in a forthcoming publication [11].

\section{Conclusion}

In conclusion, the present work shows how the concepts we have
recently introduced to exactly treat the subtle carrier exchanges which
take place in the many-body physics of excitons, can be
extended to arbitrary pairs of fermions. The correct description of
composite boson many-body effects relies on two sets of scatterings: the
``Pauli scatterings'' for fermion exchanges without interaction and the
``interaction scatterings'' for interaction without fermion exchange. To
derive these scatterings, it is not necessary for the fermion
pairs to be the exact eigenstates of the system Hamiltonian, nor to form
an orthogonal set.

\newpage

\begin{figure}[h]
\centerline{\scalebox{0.7}{\includegraphics{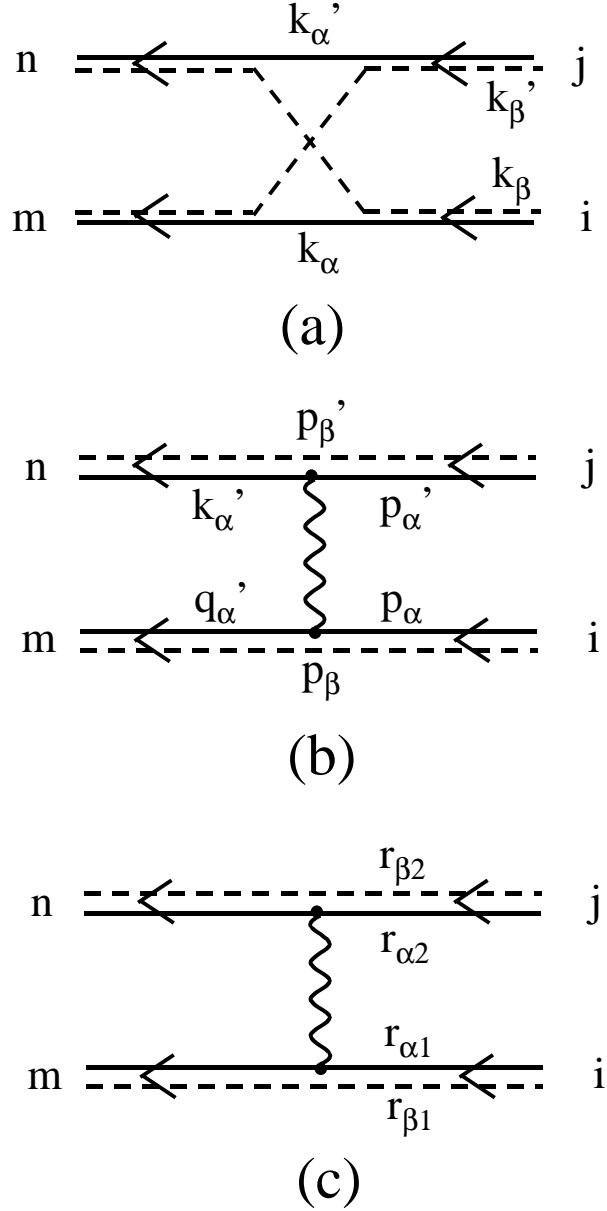}}}
\caption{(a): Diagram, in the free fermion basis $(\v k_\alpha,\v
k_\beta)$, for the ``Pauli scattering''
$\lambda\left(^{n\ \,j}_{m\ i}\right)$ given in eq.\ (16), in which the
``in'' composite bosons $i$ and $j$ exchange their fermions $\beta$,
represented by a dashed line, the ``out'' coboson $m$ being made with the
same fermion
$\alpha$ as the coboson $i$. (b): Diagram, in the free fermion basis, of
the part $\xi_1\left(^{n\ \,j}_{m\ i}\right)$, given in eq.\ (49), of the
interaction scattering, due to interactions between the fermions $\alpha$
of the ``in'' cobosons
$(i,j)$, the ``out'' cobosons $(m,n)$ being made with the same pairs as
the ``in'' cobosons. (c): Same $\xi_1\left(^{n\ \,j}_{m\
i}\right)$, due to
$(\alpha,\alpha)$ interactions, as shown in (b), but now in real
space.}
\end{figure}

\clearpage

\begin{figure}[h]
\centerline{\scalebox{0.7}{\includegraphics{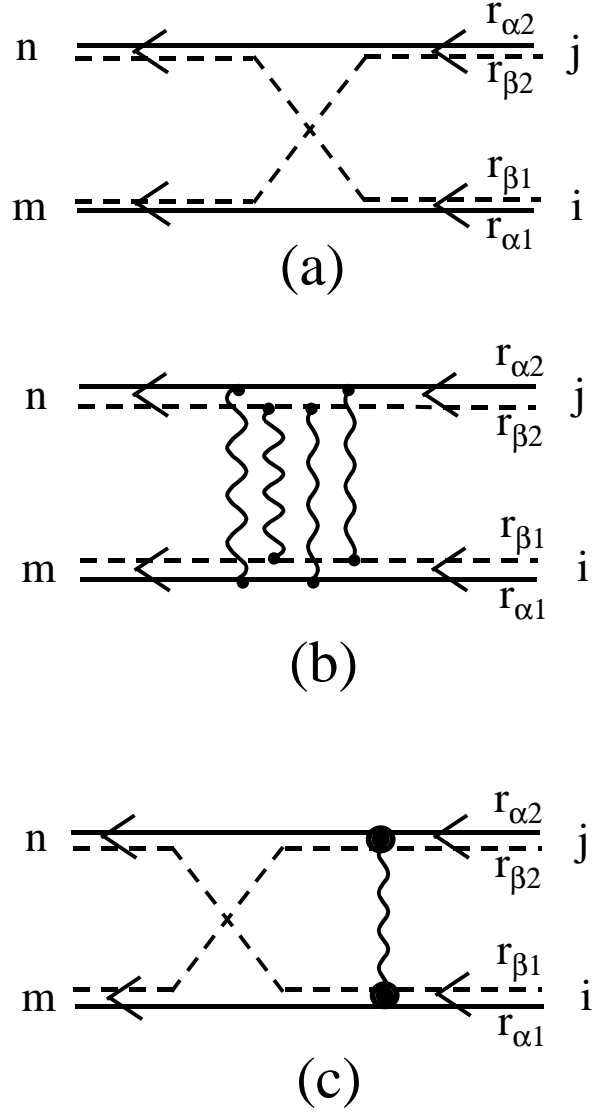}}}
\caption{(a): Diagram for the ``Pauli scattering'' $\lambda\left(^{n\
\,j}_{m\ i}\right)$, shown in Fig.1a, but now in real space. (b):
Diagram, in real space, for the ``interaction scattering''
$\xi
\left(^{n\ \,j}_{m\ i}\right)$, defined in eq.\ (51), in which the ``in''
cobosons interact through the interactions of the fermions from which
they are constructed, the ``in'' and ``out'' cobosons being made with the
same fermion pairs
$(\alpha_1,\beta_1)$ and $(\alpha_2,\beta_2)$. (c): Diagram, in real
space, for
$\xi^\mathrm{in}\left(^{n\ \,j}_{m\ i}\right)$, defined in eq.\ (54),
which is a mixed exchange-interaction scattering, the interactions taking
place before the fermion exchange, \emph{i.\ e.}, between the ``in''
cobosons $(i,j)$.}
\end{figure}

\end{document}